\documentclass{article}
\usepackage{spconf,amsmath,graphicx}
\usepackage{booktabs}
\usepackage{amssymb}
\usepackage{multirow}
\usepackage{url}
\usepackage{xcolor}

\let\OLDthebibliography\thebibliography
\renewcommand\thebibliography[1]{
  \OLDthebibliography{#1}
  \setlength{\parskip}{1pt}
  \setlength{\itemsep}{0pt plus 0.3ex}
}

\title{Speaker Adaptation for End-To-End Speech Recognition Systems in Noisy Environments}
\name{Dominik Wagner$^1$, 
Ilja Baumann$^1$, 
Sebastian P. Bayerl$^1$, 
Korbinian Riedhammer$^1$, 
Tobias Bocklet$^{1,2}$
}
\address{$^1$Technische Hochschule Nürnberg Georg Simon Ohm, Germany\\$^2$Intel Labs}
\copyrightnotice{979-8-3503-0689-7/23/\$31.00~\copyright2023 IEEE}

\begin{document}
%\ninept
%
\maketitle
%
% Motivation
% Attempts have been made to include speaker information in modern e2e architectures. 
% However this is done mostly in overlapped speech recognition.
% Features are usually added to the attention module or as a separate model
% No comparison between w2v and transformer has been done
% No ablation study under noise conditions has been done
% 
% What we do:
% We add x-vectors and ecapa features to the inputs of neural nets, in the same way it is done with hybrid HMM-DNN models, thereby training the models in a speaker-aware fashion
% The goal is to increase robustness of LVCSR systems
% We systematically show on two popular benchmarks how WER develops under different noise conditions
%
% What we show: 
% Speaker adaptation can be useful under noisy conditions even when powerful e2e architectures based on transformers and wav2vec 2.0 are used
% Effect with w2v is limited to moderate noise conditions
% effect with transformers is inverse proportional to SNR
\begin{abstract}
We analyze the impact of speaker adaptation in end-to-end automatic speech recognition models based on transformers and wav2vec 2.0 under different noise conditions. By including speaker embeddings obtained from x-vector and ECAPA-TDNN systems, as well as i-vectors, we achieve relative word error rate improvements of up to 16.3\% on LibriSpeech and up to 14.5\% on Switchboard. We show that the proven method of concatenating speaker vectors to the acoustic features and supplying them as auxiliary model inputs remains a viable option to increase the robustness of end-to-end architectures. The effect on transformer models is stronger, when more noise is added to the input speech. The most substantial benefits for systems based on wav2vec 2.0 are achieved under moderate or no noise conditions. Both x-vectors and ECAPA-TDNN embeddings outperform i-vectors as speaker representations. The optimal embedding size depends on the dataset and also varies with the noise condition.  
\end{abstract}
\section{Introduction}
\label{sec:intro}
% Embedding-Based Speaker Adaptive Training of Deep Neural Networks
% Pattern variation caused by speaker variability is one of the
% fundamental issues in acoustic modeling for automatic speech
% recognition (ASR). Such variation can give rise to covariate
% shift [1] that degrades the recognition performance. Speaker
% adaptation [2][3] and speaker adaptive training (SAT) [4][5]
% have been a common practice in ASR to reduce the covariate
% shift incurred by speaker variability.
% Speaker-aware training is also widely used for adaptation
% of DNN acoustic models, notably with i-vectors. In [11], an
% i-vector is estimated for each speaker and appended to all the
% input features from the same speaker, which is performed for
% both training and test speakers. DNNs learned from the features
% with such speaker indicators are supposed to be more speaker
% invariant
Speaker adaptation in automatic speech recognition (ASR) attempts to reduce the mismatch between training and test speakers, thereby achieving lower word error rates. 
Currently, there are two major approaches to address the speaker mismatch problem in neural network based models. 
The first approach operates in the feature space, either by normalizing acoustic features to make them speaker-independent \cite{tomashenko18,ochiai18}, or by introducing additional speaker-related knowledge (e.g. i-vectors \cite{dehak11_ivec} or x-vectors \cite{snyder18xvector}) into the acoustic model \cite{pan18_ivec,fan19_ivec}. 
% removed old refs saon13_ivec,gupta14_ivec,senior14ivec
The second approach attempts to modify the acoustic model to match the testing conditions by learning speaker or environment-dependent transformations on the inputs, outputs, or hidden representations of the neural network \cite{gemello06_adapt,liao13_adapt,meng19_adapt}. 
Supplying speaker information as an auxiliary input to acoustic models has been a popular method to achieve speaker adaptation in hybrid HMM-DNN models \cite{saon13_ivec,gupta14_ivec}.
Several studies apply the feature space approach to more recent end-to-end (E2E) architectures. 
End-to-end ASR models directly map acoustic features to a word sequence.
Popular E2E architectures based on transformers \cite{vaswani17_attention} and wav2vec 2.0 (W2V2) \cite{baevski20w2v2} have outperformed the conventional hybrid HMM-DNN framework in ASR tasks. 

Most studies on E2E systems include the speaker information in one of the lower layers of a neural network or in the attention blocks of transformer models.  
The most widely used speaker representation remain i-vectors \cite{fan19_ivec,zhao20b_interspeech,tuske21_interspeech,deng22_interspeech,zeineldeen22_interspeech,zeineldeen22improve,zhiyun19ivec}. 
In \cite{fan19_ivec}, a weighted combined speaker embedding vector is generated by applying the attention mechanism to a set of i-vectors, which helps a speech transformer model to normalize speaker variations. 
In \cite{zhao20b_interspeech}, i-vectors are concatenated with acoustic features at each self-attention layer of a speech transformer encoder, which led to relative WER improvements of 6.8\% on the Switchboard portion and 11.1\% on the Callhome portion of the Switchboard 300h Hub5'00 corpus. 
On 100 hours of LibriSpeech data, the WER is reduced by 4.5\% (\textit{test-other}) and 12.5\% (\textit{test-clean}) relative to the baseline. 
% Speech features + i-vectors form the new keys and values of the attention module
Zeineldeen et al. \cite{zeineldeen22_interspeech} propose a similar method that adds i-vectors to the input of the multi-head self-attention module in a conformer model \cite{gulati20conformer}.  
They achieve a WER reduction of 3.5\% relative to the baseline on the Callhome portion of the Switchboard 300h Hub5'00 corpus. 
In \cite{tuske21_interspeech}, i-vectors are concatenated to the input of each feedforward layer in a conformer model. 
This approach led to minor improvements in some experiments but significantly increased the number of model parameters. 

Other approaches to speaker-adaptive training for E2E architectures include x-vectors \cite{denisov19_interspeech,baskar22b_interspeech}, s-vectors \cite{shetty20}, m-vectors \cite{sari20mvector}, sequence summary networks \cite{delcroix18summarynet}, and custom speaker adaptation schemes \cite{ochiai18customadapt}. 
In \cite{shetty20}, s-vectors and x-vectors are passed through a projection layer before either adding or concatenating the output to the acoustic features in a transformer system.
Their approach achieved relative WER improvements between 7.3\% and 10.3\% on the \textit{test-clean} dataset of the LibriSpeech corpus, when x-vectors were used to supply the additional speaker information. 
However, the performance worsened between 3.6\% and 10.6\% on the \textit{test-other} dataset. 
%Appending i-vectors to the input of regular acoustic features has shown relative WER improvements on various benchmarks \cite{saon13_ivec,gupta14_ivec}
%Some works attempt to include speaker information in transformer- and conformer-based E2E systems. 
Denisov and Vu \cite{denisov19_interspeech} show that a combination of conditioning transformers on x-vectors and transfer learning improves single-channel multi-speaker speech recognition performance. 
In \cite{ochiai18customadapt}, a custom adaptation scheme for end-to-end multichannel ASR is proposed that re-estimates network parameters using target speaker speech data.
% Our work is more general in the sense that our focus lies on the evaluation of the integration of speaker embeddings into modern end-to-end speech recognizers trained for non-overlapped speech recognition. 

Existing works demonstrate the potential of speaker adaptation for E2E systems. 
However, the way speaker information is incorporated depends on the acoustic model architecture used, often makes the ASR pipeline more complex, and leads to an increase in model size. 
Furthermore, experiments are usually conducted on clean data only. 
\\
This work studies the effect of adding auxiliary speaker information to transformers and pretrained wav2vec 2.0 systems. 
Experiments on the effect of speaker adaptation for W2V2-based systems have only been conducted in dysarthric ASR \cite{baskar22b_interspeech}. 
Unlike previous works, we transfer the idea of adding speaker information to the model input of E2E architectures and study their effect under both clean and noisy conditions. 
This approach has proven successful in legacy systems and does not meaningfully increase model size and complexity. 
Furthermore, we compare the effectiveness of x-vectors \cite{snyder18xvector} and ECAPA-TDNN \cite{Desplanques2020ecapa} embeddings against an i-vector baseline. 
X-vectors have been studied less than conventional i-vectors,  and to the best of our knowledge, ECAPA-TDNN features have not yet been tested in the context of speaker adaptation, even though, they yield state-of-the-art performance in speaker recognition tasks \cite{Desplanques2020ecapa}. 
%\section{Data}
%\label{sec:data}
% 3.2. Switchboard 300
% We also evaluated the embedding-based SAT on the 300-hour
% switchboard data set. The training set consists of 262 hours of
% Switchboard 1 audio with transcripts provided by the Mississippi State University. The test set is the Hub5 2000 evaluation
% set composed of two parts: One is the 2.1-hour switchboard
% (SWB) data from 40 speakers and the other is the 1.6-hour call home (CH) data from 40 speakers.
%The training data is augmented with additive noises from the MUSAN corpus \cite{musan2015} and reverberation using a collection of room impulse responses \cite{rirs2017}. 
% \vspace{-5mm}
\section{Method}
\label{sec:method}
We conduct ASR experiments on two popular benchmarks: Switchboard \cite{swbd93} and LibriSpeech \cite{librispeech15}. 
The LibriSpeech corpus consists of 960 hours of read English speech from audiobooks sampled at 16 kHz. 
Switchboard consists of approximately 300 hours of English telephone conversations sampled at 8 kHz. 
The Switchboard audio is resampled to 16 kHz for the experiments involving the W2V2-based system, to make it compatible with the pretrained feature extractor. 
The speaker recognition systems (i-vector, x-vector, and ECAPA-TDNN) used to generate auxiliary inputs, are trained on the VoxCeleb \cite{Nagrani17} dataset, which is comprised of 1.2 million utterances from 7.3k different speakers. 

Baseline transformer- and W2V2-based ASR models are trained with minor architectural differences depending on the dataset. 
The same models are trained with additional speaker embeddings included at different parts of the overall system. 
%TODO: Add contribution again at some point
% The recipes for training the baseline models based on wav2vec 2.0 and transformers on Switchboard data to the SpeechBrain project 
The block diagrams in Figure \ref{fig:architecture_w2v} and Figure \ref{fig:architecture_trafo} illustrate the composition of both architectures used in this work, as well as the integration of speaker embeddings. 
Parameters that are kept constant throughout all experiments (e.g. dropout probability or channels in convolutional layers) are included in the figures. 

We use 80-dimensional Mel filterbank features with a frame-width of 25ms and a frame-shift of 10ms as acoustic input features for the transformer, the ECAPA-TDNN and the x-vector system. 
The pretrained wav2vec 2.0 model receives raw waveform inputs. 
We employ 1000 subword units \cite{kudo18entencepiece} estimated via unigram language modeling as recognition tokens for the ASR models. 
To evaluate the effect of speaker adaptation under varying noise conditions, the test data is augmented with randomly chosen additive noises from the MUSAN corpus \cite{musan2015} at constant signal-to-noise ratios of 18, 9, and 0. 
We choose the speech portion of the MUSAN corpus, which comprises approximately 60 hours of various speech recordings. 
%TODO: Maybe add upsampling of swbd to 16khz
\subsection{Pretrained wav2vec 2.0}
\begin{figure}[t]
  \centering
    \includegraphics[width=0.99\linewidth]{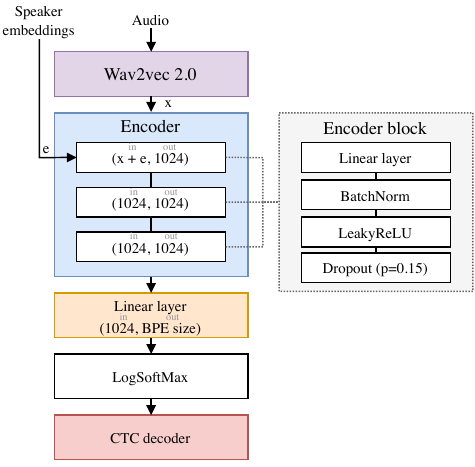}
      \vspace{-4mm}
    \caption{Pretrained wav2vec 2.0 system and integration of speaker embeddings.}
  \label{fig:architecture_w2v}
  % \vspace{-7.5mm}
\end{figure}
The pretrained wav2vec 2.0 model used for acoustic feature extraction is described in \cite{baevski20w2v2}. 
It consists of a convolutional feature encoder with multiple identical blocks using temporal convolution, layer normalization, and a GELU activation function, followed by 12 transformer encoder blocks. 
The convolutional feature encoder generates a sequence of embeddings for each utterance, which are then passed to the transformer encoder to capture information about the entire input sequence. 
For self-supervised training, the output of the convolutional feature encoder is discretized to a finite set of speech representations using product quantization. 
We use a model pretrained and finetuned on 960 hours of Libri-Light \cite{kahn20librilight} and LibriSpeech \cite{librispeech15} data. 

%Pretraining was conducted with the audio data of the LibriSpeech corpus without transcriptions and the audio data from the LibriVox project resulting in 53.2k hours of audio.
% Model was trained with Self-Training objective. 
% torch.Size([8, 89, 1024])
% 50 frames for 1 sec of audio
The model components are illustrated in Figure \ref{fig:architecture_w2v}. 
The W2V2 feature extractor generates approximately 50 acoustic representations in $\mathbb{R}^{1024}$ for 1 second of raw audio input. 
In our system, these acoustic features are concatenated with the speaker vectors and passed through three encoder blocks consisting of a linear layer, batch normalization \cite{ioffe15batchnorm}, leakyReLU activation and dropout \cite{srivastava14droput}. 
A final linear layer yields outputs of vocabulary size (1000 subword units). 

Since the wav2vec 2.0 model has already been pretrained on LibriSpeech, we keep its weights frozen during training on the LibriSpeech corpus, but include them in the training procedure, when the Switchboard data is used. 

The models are trained with SpecAugment \cite{specaugment2019} in the time-domain and speed perturbation at 95\% and 105\% of the regular utterance speed. 
Each model is trained for 30 epochs on Switchboard and for 5 epochs on LibriSpeech. 

The batch size is 6 for both datasets. 
We employ the Adam \cite{kingma14_adam} optimizer for the W2V2 part of the overall system with exponential decay rates of $\beta_1 = 0.90$, $\beta_2 = 0.99$ and an initial learning rate of $10^{-3}$. 
The other parts of the model are trained using the Adadelta \cite{zeiler12adadelta} optimizer with an initial learning rate of $\alpha=1.0$ and decay rate of $\rho = 0.95$. 
The learning rate is annealed based on the validation performance. 

Greedy decoding on the logits and application of the connectionist temporal classification (CTC) rules (removal of blank symbols and duplicates) yields the best path for each utterance. 
% \vspace{-5mm}
\subsection{Transformer}
\begin{figure}[t]
  \centering
    \includegraphics[width=0.97\linewidth]{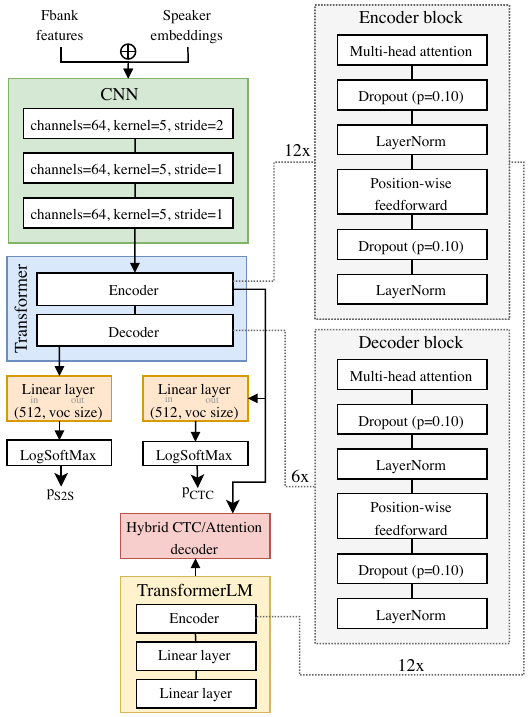}
      \vspace{-3mm}
    \caption{Transformer and speaker embedding integration.}
  \label{fig:architecture_trafo}
  \vspace{-5mm}
\end{figure}
We employ an attention-based encoder-decoder architecture \cite{watanabe17_ctcatt}. 
%In particular, CTC+Att systems rely on an encoder-decoder architecture with an additional CTC loss applied on the top of the encoder
The model is trained on both CTC loss and Kullback-Leibler divergence between negative-log likelihood targets. 
Decoding is performed with a joint CTC/attention beam search coupled with a transformer language model that is used on top of the decoder probabilities. 

The composition of the transformer system is depicted in Figure \ref{fig:architecture_trafo}. 
For each input sequence, a convolutional neural network (CNN) consisting of 3 layers is applied to reduce the length of the hidden representations before passing the input to the transformer model. 
The transformer architecture follows \cite{vaswani17_attention}. 
It employs 12 identical encoder blocks and 6 identical decoder blocks with 4 heads in the multi-head attention modules. 
Information about the token position is encoded via sinusoidal positional encodings in 256-dimensional input embeddings. 
Each encoder and decoder block consists of two sub-layers; a multi-head attention mechanism and a position-wise fully connected feedforward network. 
Dropout and layer normalization \cite{ba16layernorm} are applied after each sub-layer. 

The transformer-encoder output is passed to the hybrid CTC/attention decoder and to a linear layer followed by log softmax activation. 
The output after log softmax activation $p_{CTC}$ is used to compute the CTC loss.  %and the best path in the hybrid CTC/attention beam search. 
The transformer-decoder output is passed through another linear layer followed by log softmax activation. 
The resulting log-probabilities $p_{S2S}$ are used to compute the Kullback-Leibler loss. 

%Swbd
% lm_model: !new:speechbrain.lobes.models.transformer.TransformerLM.TransformerLM # yamllint disable-line rule:line-length
%   vocab: !ref <output_neurons>
%   d_model: 264 Embedding dimension size.
%   d_embedding: 128
%   nhead: 12
%   num_encoder_layers: 12
%   num_decoder_layers: 0
%   d_ffn: 1024
%   dropout: 0.1
%   activation: !name:torch.nn.ReLU
%   normalize_before: False
The acoustic model is coupled with a transformer language model (LM) during decoding. 
Its architecture is also based on \cite{vaswani17_attention}. 
The LM employs 12 encoder blocks but no decoder blocks. 
Each block in the LibriSpeech model employs 768-dimensional input embeddings, the hidden layer size is 3072 and 12 heads are used in the multi-head attention module. 
The encoder blocks in the Switchboard model use 264-dimensional input embeddings, a hidden layer size of 1024 and 12 attention heads. 

The LibriSpeech tokenizer and LM are trained on 960h of audio transcripts, which corresponds to 10 million words of text data. 
The Switchboard tokenizer and LM are trained on the transcripts of the Switchboard 300h corpus, as well as the transcripts provided by the Fisher corpus \cite{Cieri2004TheFC}. 
The LM weight is 0.6 for the LibriSpeech models and 0.3 for the Switchboard models. 
The beam size of the decoder is 60 for both datasets. 

The LibriSpeech transformer is trained for 60 epochs with an effective batch size of 48 using gradient accumulation with a factor of 3 and a CTC weight of 0.3. 
The Switchboard transformer is trained for 100 epochs with an effective batch size of 256 using gradient accumulation with a factor of 2 and a CTC weight of 0.3. 
We employ the Adam optimizer with exponential decay rates of $\beta_1 = 0.90$, $\beta_2 = 0.98$ and initial learning rates of $10^{-3}$ (LibriSpeech) and $6\times10^{-3}$ (Switchboard). 
The learning rate is increased linearly for the first 25k steps and then decreased proportionally to the inverse square root of the step number \cite{vaswani17_attention}. 
% \vspace{-2mm}
\subsection{Speaker embeddings}
% \vspace{-1mm}
Our i-vector extraction pipeline follows Kaldi's \cite{povey11kaldi} \texttt{sre10/v2} recipe, using a 2048-component diagonal universal background model (UBM) trained on 20-dimensional MFCC features. 

We utilize two state-of-the-art systems to extract speaker embeddings: x-vector \cite{snyder18xvector} and ECAPA-TDNN \cite{Desplanques2020ecapa}. 
The x-vector model \cite{snyder18xvector} is a time delay neural network (TDNN) that aggregates variable-length inputs across time via mean pooling to create fixed-length representations capable of capturing speaker characteristics. 
Speaker embeddings are extracted from a bottleneck layer prior to the output layer. 
%The model uses the time delay neural network (TDNN) architecture and consists of nine layers. 
% The first five layers operate on frame level with three layers having a left and right temporal context of $\lbrace 2,2,3 \rbrace$ and 512 nodes followed by two layers with 512 and 1500 nodes seeing a total context of 15 frames each.
%The first five layers of the model work on frame-level with a small context in the time domain (2 or 3 frames) to the left and right of the current frame.
%The input to the next layer is the spliced output of the current frame and the output of the previous layer at the current frame plus/minus the context. 
%The sixth layer aggregates its inputs across time using mean and standard deviation to create fixed-length outputs (statistics pooling).  
%The statistics pooling layer generates fixed-length outputs using mean and standard deviation.  
%The two remaining layers operate on the aggregated information. 
%A final softmax layer is sized according to the number of speakers in the corpus. 
%Nonlinearities across the layers are introduced via rectified linear units (ReLUs). 
%The model is trained to discriminate between the speakers in the training data. 
%Each training example consists of MFCC features representing a chunk of audio with a length between 2 seconds and 4 seconds as well as a speaker label. 

Desplanques et al. \cite{Desplanques2020ecapa} propose several enhancements to the x-vector architecture. 
Their ECAPA-TDNN adds 1-dimensional Res2Net \cite{gao2021res2net} modules with skip connections as well as squeeze-excitation (SE) blocks \cite{jie2018squeeze} to capture channel interdependencies. 
Additionally, features are aggregated and propagated across multiple layers. 
It also utilizes a channel-dependent self-attention mechanism that uses a global context in the frame-level layers and the statistics pooling layer. 
%It aims to capture the importance of each frame given the channel and is used to compute a weighted mean and standard deviation for the channel. 
%The final output of the pooling layer is a concatenation of the channel-wise weighted mean and standard deviation vectors.  
%The activations of the last frame-level layer at a certain time step are concatenated with the global non-weighted mean and standard deviation of this layer across the time domain. 
%This allows the attention mechanism to take global properties of the utterance into account. 
% The main purpose of SE blocks is the expansion of the temporal context of a frame-level layer by rescaling the channels according to global properties of the recording. 
% Frame-level features are rescaled to reflect a more global context with 1-dimensional SE blocks. 
% SE blocks consist of a squeeze operation and a excitation operation. 
% The squeeze operation consists of calculating the mean vector of the frame-level features across the time domain for each channel. 
% The means are then used in the excitation operation to calculate a weight for each channel. 
% The squeeze operation acts as a bottleneck layer. 
% Its result contains weights that scale the original input via channel-wise multiplication.
% SB: too much detail, its an SE Block -> Ref to SE paper enough
%We use the ECAPA-TDNN implementation from \cite{speechbrain}. 
The input features are 24-dimensional Mel filterbank features using a frame-width of 25ms and a frame-shift of 10ms. 
% Additionally, the data is speed-perturbed at 95\% and 105\% of the normal utterance speed and SpecAugment \cite{specaugment2019} is applied in the time domain. 

The training data for both the x-vector model and the ECAPA-TDNN are augmented with additive noises from the MUSAN corpus \cite{musan2015} and reverberation using a collection of room impulse responses. 

To ensure a fair comparison, we extract 512-dimensional i-vectors, x-vectors and ECAPA-TDNN embeddings for each utterance in the training and test set. 
We also trained an ECAPA-TDNN that yields 192-dimensional outputs in the final fully-connected layer, as proposed in the original paper \cite{Desplanques2020ecapa}, to analyze the impact of reduced dimensionality. 
We compute the mean over all vectors representing utterances from an individual speaker, thereby generating a single prototype embedding for each speaker in the corpus. 
The mean vectors are concatenated with the acoustic features and passed to the ASR systems. 
Each embedding vector is scaled to the value range $[0;1]$. 
We also experimented with unscaled vectors and scaling to zero mean and unit variance. 
However, the former yielded the most consistent results. 
% \vspace{-4mm}
\section{Experiments}
\label{sec:experiments}
% \vspace{-1mm}
\setlength{\tabcolsep}{1pt} % Default value: 6pt
\renewcommand{\arraystretch}{1} % Default value: 1
\begin{table*}
\caption{Word error rates (WER) and relative improvements for the pretrained W2V2 (left) and the transformer (right) model under varying noise conditions. 
The column $\Delta$\% indicates the improvement of speaker adaptation over the baseline in percent. 
Bold values indicate the largest relative WER improvements for the respective test set. 
The leftmost column shows the signal-to-noise ratio (SNR) used in the experiment. 
The test sets \textit{clean} and \textit{other} represent the ``test-clean'' and ``test-other'' sets of the LibriSpeech corpus. 
\textit{SW}, \textit{CH}, and \textit{Hub'00} indicate the Switchboard, Callhome, and full Hub5'00 portions of the Switchboard 300h Hub5'00 dataset. 
}
\label{tab:results}
% \vspace{-2mm}
\scalebox{0.72}{
\centering
\begin{tabular}[t]{cccc|cc|cc|cc|cc}
\multicolumn{12}{c}{\normalsize{\textbf{Pretrained wav2vec 2.0}}}\\
\toprule
\textbf{SNR}        & \textbf{Corpus}        & \textbf{Data} & \textbf{Baseline} & \multicolumn{2}{c|}{\textbf{ECAPA} $\mathbb{R}^{192}$} & \multicolumn{2}{c|}{\textbf{ECAPA} $\mathbb{R}^{512}$} & \multicolumn{2}{c|}{\textbf{X-vector} }  & \multicolumn{2}{c}{\textbf{i-vector}}\\
         &            &   & \footnotesize{\textit{WER}  }    & \footnotesize{\textit{WER}}          & \footnotesize{$\Delta$\% }  & \footnotesize{\textit{WER}}          & \footnotesize{$\Delta$\% }           & \footnotesize{\textit{WER}}         & \footnotesize{$\Delta$\%  }       & \footnotesize{\textit{WER}}          & \footnotesize{$\Delta$\% }    \\
\midrule
\multirow{5}{*}{-}  & \multirow{2}{*}{Libri} & clean  & 1.89 & 1.88    & 0.5\% & 1.89 & 0.0\%          & 1.89 & 0.0\%       & 1.92 & -1.6\%   \\
                    &                        & other  & 3.99 & 3.97    & 0.5\%  & 4.03 & -1.0\%         & 3.95 & 1.0\%      & 3.95 & 1.0\%  \\\cline{2-12}
                    & \multirow{3}{*}{Swbd}  & SW     & 8.78 & 8.60    & 2.1\%  & 8.33 & 5.1\%         & 8.31 & 5.4\%            & 8.46  & 3.6\%  \\
                    &                        & CH     & 15.19 & 14.98  & 1.4\%  & 14.94 & 1.6\%         & 14.23 & \textbf{6.3\%} & 15.23  & -0.3\% \\
                    &                        & Hub'00 & 11.90 & 11.71   & 1.6\% & 11.63 & 2.3\%          & 11.22& \textbf{5.7\%} & 11.81 & 0.8\%  \\
\hline
\multirow{5}{*}{18} & \multirow{2}{*}{Libri} & clean & 9.27  & 8.91  & 3.9\%    & 8.72 & \textbf{5.9}\%          & 8.92  & 3.8\%    &  8.93 & 3.7\%        \\
                    &                        & other & 18.89 & 18.45 & \textbf{2.3\%} & 18.88 & 0.1\%    & 18.61 & 1.5\%   & 19.20 & -1.6\%       \\\cline{2-12}
                    & \multirow{3}{*}{Swbd}  & SW    & 9.52  & 9.45  & 0.7\%   & 9.12 & 4.2\%           & 8.99  & \textbf{5.6\%} & 9.37 & 1.6\%  \\
                    &                        & CH    & 18.07 & 17.89 & 1.0\%   & 18.36 & -1.6\%           & 17.41 & 3.7\%   & 18.24 & -0.9\%       \\
                    &                        & Hub'00& 13.62 & 13.56 & 0.4\%   & 13.29 & 2.4\%           & 13.22 & 2.9\%    & 13.79 & -1.2\%      \\
\hline
\multirow{5}{*}{9}  & \multirow{2}{*}{Libri} & clean & 21.43 & 21.21 & 1.0\%   & 20.49 & 4.4\%        & 20.55 & 4.1\% & 21.13 & 1.4\% \\
                    &                        & other & 37.79 & 37.93 & -0.4\%   & 38.24 & -1.2\%        & 37.17 & 1.6\%    & 38.73 & -2.5\%      \\\cline{2-12}
                    & \multirow{3}{*}{Swbd}  & SW    & 11.95 & 12.08 & -1.1\%   & 11.34 & 5.1\%       & 11.73 & 1.8\%     & 11.96 & -0.1\%     \\
                    &                        & CH    & 26.62 & 26.58 & 0.2\%    & 27.74 & -4.2\%       & 27.62 & -3.8\%   & 28.30 & -6.3\%      \\
                    &                        & Hub'00& 19.88 & 19.84 & 0.2\%    & 19.84 & 0.2\%        & 20.02 & -0.7\%   & 19.66 & 1.1\%      \\
\hline
\multirow{5}{*}{0}  & \multirow{2}{*}{Libri} & clean  & 41.26  & 41.19 & 0.2\%  & 41.92 & -1.5\%         & 40.85 & 1.0\%   & 43.25 & -4.8\%       \\
                    &                        & other  & 62.24 & 61.94  & 0.5\%  & 62.50 & -0.4\%         & 61.21 & 1.7\%   & 64.74 & -4.0\%       \\\cline{2-12}
                    & \multirow{3}{*}{Swbd}  & SW    & 21.71  & 23.04  & -6.1\% & 22.55 & -3.9\%         & 22.23 & -2.4\%   & 21.49 & 1.0\%      \\
                    &                        & CH     & 46.86 & 48.80  & -4.1\% & 50.24 & -7.2\%         & 49.81 & -6.3\%     & 50.40 & -7.5\%    \\
                    &                        & Hub'00 & 35.38 & 35.90  & -1.5\% & 35.86 & -1.4\%         & 36.34  & -2.7\% & 35.34 & 0.1\%        \\ 
\bottomrule
\end{tabular}
}
\hfill%
\scalebox{0.72}{
\begin{tabular}[t]{cccc|cc|cc|cc|cc}
\multicolumn{12}{c}{\normalsize{\textbf{Transformer}}}\\
\toprule
\textbf{SNR}        & \textbf{Corpus}        & \textbf{Data} & \textbf{Baseline} & \multicolumn{2}{c|}{\textbf{ECAPA} $\mathbb{R}^{192}$} & \multicolumn{2}{c|}{\textbf{ECAPA} $\mathbb{R}^{512}$} & \multicolumn{2}{c|}{\textbf{X-vector}}  & \multicolumn{2}{c}{\textbf{i-vector}}\\
         &            &   & \footnotesize{\textit{WER}  }    & \footnotesize{\textit{WER}}          & \footnotesize{$\Delta$\% }  & \footnotesize{\textit{WER}}          & \footnotesize{$\Delta$\% }           & \footnotesize{\textit{WER}}         & \footnotesize{$\Delta$\%  }       & \footnotesize{\textit{WER}}          & \footnotesize{$\Delta$\% }    \\
\midrule
\multirow{5}{*}{-}  & \multirow{2}{*}{Libri} & clean     & 2.39    & 2.34    & 2.1\%    & 2.36 & 1.3\%     & 2.31    & 3.3\%   & 2.43 & -1.7\%       \\
                    &                        & other     & 5.42    & 5.69    & -5.0\%   & 5.85 & -7.9\%     & 5.57     & -2.8\% & 6.03 & -11.3\%      \\\cline{2-12}
                    & \multirow{3}{*}{Swbd}  & SW        & 10.37   & 10.22   & 1.4\%    & 9.99 &  3.7\%     & 10.09    & 2.7\%   & 10.32 & 0.5\%       \\
                    &                        & CH        & 18.52   & 18.22   & 1.6\%    & 18.88 & -1.9\%   & 18.59      & -0.4\%  & 20.07 & -8.4\%       \\
                    &                        & Hub'00    & 14.44   & 14.17   & 1.9\%    & 14.63 & -1.3\%     & 14.62    & -1.2\%  & 15.29 & -5.9\%       \\
\hline
\multirow{5}{*}{18} & \multirow{2}{*}{Libri} & clean     & 9.95    & 9.26   & 6.9\%     & 8.33 & \textbf{16.3}\%     & 9.17     & 7.8\%     & 8.47 & 14.9\%   \\
                    &                        & other     & 22.11   & 19.99  & 9.6\% & 19.55 & \textbf{11.6\%} & 20.38 & 7.8\%  & 20.79 & 6.0\%     \\\cline{2-12}
                    & \multirow{3}{*}{Swbd}  & SW        & 13.72   & 13.38  & 2.5\%     & 12.71 & 7.4\%      & 13.01   & 5.2\%  & 13.65 & 0.5\%       \\
                    &                        & CH        & 26.75   & 25.97  & 2.9\%     & 25.24 & 5.6\%      & 25.34   & 5.3\%  & 26.72 & 0.1\%        \\
                    &                        & Hub'00    & 20.37   & 19.41  & 4.7\%     & 19.43 & 4.6\%      & 19.47   & 4.4\%  & 20.28 & 0.4\%        \\
\hline
\multirow{5}{*}{9}  & \multirow{2}{*}{Libri} & clean    & 25.29   & 25.27   & 0.1\%     & 22.79 & 9.9\%      & 23.36   & 7.6\% & 23.21 & 8.2\%         \\
                    &                        & other    & 45.27   & 43.26   & 4.4\%     & 41.08 & 9.3\%      & 41.97    & 7.3\% & 42.75 & 5.6\%         \\\cline{2-12}
                    & \multirow{3}{*}{Swbd}  & SW       & 20.03   & 20.03   & 0.0\%     & 19.53 & 2.5\%      & 18.79    & 6.2\%  & 20.84 & -4.0\%        \\
                    &                        & CH       & 39.59   & 35.08   & 11.4\%    & 35.18 & 11.1\%      & 34.96    & 11.7\% & 35.82 & 9.5\%        \\
                    &                        & Hub'00   & 30.88   & 26.99   &\textbf{12.6\%}& 27.91 & 9.6\%   & 27.24  & 11.8\%  & 28.54 & 7.6\%       \\
\hline
\multirow{5}{*}{0}  & \multirow{2}{*}{Libri} & clean    & 51.87   & 48.94   & 5.6\%    & 46.95 & 9.5\%        & 47.53 & 8.4\%   & 46.85 & 9.7\%      \\
                    &                        & other    & 71.90   & 68.80   & 4.3\%    & 66.88 & 7.0\%       & 68.54  & 4.7\%    & 69.45 & 3.4\% \\\cline{2-12}
                    & \multirow{3}{*}{Swbd}  & SW       & 35.84   & 33.12   & 7.6\%    & 33.51 & 6.5\%      & 31.32   & \textbf{12.6\%}  & 36.32 & -1.3\%     \\
                    &                        & CH       & 56.15   & 48.78   & 13.1\%   & 50.02 & 10.9\%      & 48.03    & \textbf{14.5\%} & 50.91 & 9.3\%    \\
                    &                        & Hub'00   & 45.83   & 40.74   & 11.1\%   & 41.82 & 8.7\%      & 40.43   & 11.8\%          & 44.03 & 3.9\%\\   
\bottomrule
\end{tabular}
}
%\vspace{-1.8mm}
% \vspace{-5mm}
\end{table*}
We trained five models for each corpus using both the W2V2 and the transformer architecture: (1) model without speaker adaptation, (2) i-vectors as auxiliary inputs, (3) 192-dimensional ECAPA-TDNN embeddings as auxiliary inputs, (4) 512-dimensional ECAPA-TDNN embeddings as auxiliary inputs, (5) x-vectors as auxiliary inputs. 
Each system was evaluated using clean and augmented (SNRs 18, 9, and 0) test data. 

The left part of Table \ref{tab:results} shows the word error rates (WERs) and improvements relative to the baseline without speaker adaptation for the W2V2-based models. 
WER improvements were achieved with both x-vectors and ECAPA-TDNN embeddings under all conditions, except at $SNR=0$. 
X-vectors and ECAPA-TDNN embeddings outperformed i-vectors across almost all conditions and datasets.  
X-vectors achieved the largest WER improvements on the Switchboard data, whereas ECAPA-TDNN embeddings were more effective on the LibriSpeech data. 
Performance gains diminished or turned negative for SNRs 9 and 0 across all types of embeddings. 
The largest overall improvement of 6.3\% was achieved on the Callhome portion of Switchboard, when no noise augmentation was applied to the data and x-vectors were used as speaker information. 
%Word error rate improvements started to turn negative for both ECAPA-TDNN and x-vector adaptation at SNRs below 18.  

The right part of Table \ref{tab:results} shows the ASR results for the transformer models. 
WER improvements had the tendency to increase with smaller SNRs, i.e., larger performance gains were achieved, when more noise was added to the test data. 
%X-vectors often outperformed ECAPA-TDNN embeddings as auxiliary features in terms of relative WER changes under strong noise conditions ($SNR\geq9$). 
The largest overall improvement of 16.3\% was achieved on the clean test portion of the LibriSpeech corpus ($SNR=18$) with 512-dimensional ECAPA-TDNN embeddings as auxiliary features. 
X-vectors and ECAPA-TDNN embeddings outperformed i-vectors across all noise conditions, except one (LibriSpeech \textit{test-clean}, $SNR=0$). 
%I-vectors achieved the second largest overall improvement (14.9\%) on the clean portion of LibriSpeech at an SNR of 18. 
%The results for ECAPA-TDNN embeddings and x-vector embeddings were inconsistent, when no noise was added to the testdata. 
%For example, speaker adaptation with the x-vector system yielded negative improvement rates for all three Switchboard test sets and the \textit{test-other} set of LibriSpeech, whereas adaptation with ECAPA-TDNN features only turned negative on the \textit{test-other} set of LibriSpeech.
\\ 
A direct comparison between the left and the right part of Table \ref{tab:results} shows that the W2V2-based models yielded consistently lower WERs than the transformer systems, but the relative WER gains due to speaker adaptation were larger for the transformer models. 
\\
For comparison, we trained hybrid HMM-DNN systems using the lattice-free maximum mutual information (LF-MMI) objective \cite{povey16_interspeech} on LibriSpeech and Switchboard with the \texttt{s5} and \texttt{s5c} recipes of the Kaldi toolkit. 
On LibriSpeech, we achieved relative improvements over the baseline WER of 8.8\% on \textit{test-clean} and 13.9\% on \textit{test-other}, when i-vectors were used as auxiliary inputs to the acoustic model and no noise was added to the test data. 
On Switchboard, improvements of 13.5\% (\textit{SW}) and 15.6\% (\textit{CH}) were reached. 
%Combining speaker information with acoustic features is a proven method in hybrid HMM-DNN speech recognition. 
The transformer models yielded only small and rather inconsistent improvements in terms of WER on clean test data (cf. first five rows in right part of Table \ref{tab:results}). 
However, under strong noise conditions ($SNR=0$), our experiments showed performance gains comparable to those of HMM-DNN systems. 
Our W2V2-based approach showed more consistent improvements on clean data and moderate noise ($SNR=18$) than the transformer approach, but did not yield gains as high as the HMM-TDNN systems.  
% \vspace{-3.5mm}
\section{Discussion}
\label{sec:discussion}
% \vspace{-1mm}
The overall performance gains of speaker adaptation on the W2V2-based systems were smaller than those on the transformer systems. 
We assume that this is in parts related to the better overall performance of finetuned wav2vec 2.0 systems and to the stage at which the speaker information was included in the model. 
Speaker information was added at the first encoder block of the W2V2-based system, but already included in the convolutional frontend of the transformer models (cf. Figure \ref{fig:architecture_trafo}). 
In other words, speaker embedding inclusion in the W2V2-based system happened in a late stage of the model, at which most of the acoustic information was already processed by the feature extractor, whereas in the transformer system speaker embeddings were added in an early stage, enabling it to jointly model acoustics and speaker information throughout all layers.
In fact, when we included speaker embeddings at the linear layers after the transformer component, the benefits shown in the right part of Table \ref{tab:results} disappeared.  
Furthermore, wav2vec 2.0 systems are known to capture speaker information well \cite{fan21_interspeech}, consequently making the integration of additional speaker embeddings in the downstream encoder layers less important.  
%W2V2 does not yield consistent improvements since we inject the speaker embeddings at the first of the two CNN layers, after the W2V2 features had already been extracted. 
%Passing audio through the wav2vec 2.0 model is merely a feature extraction step, which models all of the acoustics. 
%Hence, there is not much to be learned in the final two layers we added. 
%The transformer system on the hand receives the embeddings in an early stage, where acoustics are still modeled. 
%Our experiments yielded smaller WER improvements when ECAPA-TDNN embeddings were used instead of x-vectors. 
\\
The experiments on ECAPA-TDNN embeddings with different sizes showed that dimensionality can affect the overall performance, but it seems to be linked to other factors such as model architecture and noise condition as well.  
In most cases, using transformers with ECAPA-TDNN embeddings in $\mathbb{R}^{512}$, instead of $\mathbb{R}^{192}$, yielded larger WER improvements. 
However, for the W2V2-based system, lower dimensional ECAPA-TDNN embeddings were more beneficial than higher dimensional ones under noisy environments. 
%We assume that this might be related to the lower dimensionality ($\mathbb{R}^{192}$ instead of $\mathbb{R}^{512}$) of the features. 
The choice of embedding extractor also depends on the speech corpus. 
Our experiments showed that the largest overall gains on Switchboard were achieved with x-vectors, whereas the largest gains on LibriSpeech were achieved using ECAPA-TDNN embeddings. 
% \setlength{\tabcolsep}{2pt} % Default value: 6pt
% \renewcommand{\arraystretch}{1} % Default value: 1
% \begin{table}
% \caption{Word error rates (WER) and relative improvements for hybrid HMM-DNN systems.}
% \label{tab:results_kaldi}
% \centering
% \scalebox{0.95}{
% % Please add the following required packages to your document preamble:
% % \usepackage{multirow}
% \begin{tabular*}{0.5\textwidth}{c @{\extracolsep{\fill}}ccc|cc}
% \toprule
% \textbf{SNR}        & \textbf{Corpus}        & \textbf{test set} & \textbf{Baseline} & \multicolumn{2}{c}{\textbf{i-vector}}     \\
%          &            &   & \footnotesize{\textit{WER}  }    & \footnotesize{\textit{WER}}          & \footnotesize{$\Delta$\% }  \\
% \midrule
% \multirow{5}{*}{-} & \multirow{2}{*}{Libri} & clean            & 5.48              & 5.00                 & 8.8\%          \\
%                   &                        & other            & 14.58             & 12.56                & 13.9\%         \\\cline{2-6}
%                   & \multirow{3}{*}{Swbd}  & SW               & 14.10             & 12.20                & 13.5\%         \\
%                   &                        & CH               & 27.60             & 23.30                & 15.6\%         \\
%                   &                        & Hub'00           & 20.90             & 17.80                & 14.8\%         \\
% \bottomrule
% \end{tabular*}
% }
% \end{table}
% \vspace{-3.5mm}
\section{Conclusion}
\label{sec:conclusion}
% \vspace{-1mm}
We analyzed the impact of speaker adaptation on ASR performance using two recent E2E architectures.
We demonstrated that the use of x-vectors and ECAPA-TDNN embeddings as auxiliary input features can increase the robustness in noisy environments. 
Both representation types outperformed i-vectors in almost all cases. 
The system based on wav2vec 2.0 yielded consistent improvements with speaker adaptive training, when no or small amounts of noise were added ($SNR\geq18$), but did not achieve improvements under strong noise conditions ($SNR\leq9$). 
Transformers achieved larger WER improvements, the more heavily the test data was augmented. 
%The largest performance gains for both W2V2-based and transformer models were achieved using x-vectors. 
%However, ECAPA-TDNN embeddings are more efficient due to their lower dimensionality and therefore remain an option to consider in practical usage. 
Both E2E architectures can benefit from additional speaker information, but the expected performance gains are lower compared to hybrid HMM-DNN systems and depend on the dataset, the strength of the noises applied, as well as the type of speaker representations used. 
Future work will extend this analysis to other E2E models such as conformers to complete the picture on the impact of speaker embeddings as auxiliary input features. 
% References should be produced using the bibtex program from suitable
% BiBTeX files (here: strings, refs, manuals). The IEEEbib.bst bibliography
% style file from IEEE produces unsorted bibliography list.
% -------------------------------------------------------------------------
\newpage
\bibliographystyle{IEEEbib}
\footnotesize{
\bibliography{strings,refs}
}
\end{document}